\documentstyle[epsf,epsfig,preprint,aps,amssymb]{revtex}
\draft \preprint{KIAS-P01065; SNUTP 01/041}
\begin{document}
\title{\Large\bf
 Top-bottom mass hierarchy, $s-\mu$ puzzle and gauge coupling
 unification with split multiplets}
\author{$^{(a)}$Hyung Do Kim\footnote{hdkim@kias.re.kr},
$^{(b)}$Jihn E. Kim\footnote{jekim@phyp.snu.ac.kr} and
$^{(b)}$Hyun Min Lee\footnote{minlee@phya.snu.ac.kr}}
\address{$^{(a)}$School of Physics, Korea Institute for Advanced Study, Cheongryangri-dong,
Dongdaemun-ku, Seoul 135-012, Korea\\
$^{(b)}$School of Physics and Center for Theoretical Physics,
Seoul National University, Seoul 151-747, Korea} \maketitle

\begin{abstract}
A supersymmetric 5D $SU(5)$ grand unification is considered. The
$SU(5)$ is broken down to $G_{SM}=SU(3)\times SU(2)\times U(1)$ by
the $Z_2\times Z_2'$ assignment of the bulk field(s). The matter
fields are located at the fixed point(s). In the bulk, a Higgs
multiplet $\bar 5_H$(containing the bottom doublet $H_1$) and the
$SU(5)$ gauge multiplet are located. At one fixed point, $H_2$(the
top doublet) and the standard model matter multiplets are
presented. Because of the difference of the locations of $H_1$ and
$H_2$, one can obtain a hierarchy between top and bottom Yukawa
couplings. We also present a possibility to understand the $s-\mu$
mass puzzle in this framework of the split multiplet.
\\
\vskip 0.5cm\noindent [Key words: split multiplet, top-bottom mass
hierarchy, $s-\mu$ puzzle]
\end{abstract}


\newpage

\section{Introduction}

The unification of gauge coupling constants is an attractive
proposal under the name of grand unification\cite{georgi}, which
cannot be understood in the standard model(SM). At the unification
scale $M_U$ the strong, weak and electromagnetic coupling
constants are the same since they are described by a simple or
semi-simple group $G$ for grand unification. Below the unification
scale, this grand unification(GUT) group $G$ is broken down to the
standard model group $SU(3)\times SU(2)\times U(1)$ and the
difference between the SM couplings is generated\cite{gqw}. Since
the GUT unifies the interactions and some quarks and leptons are
assigned in a same $G$ multiplet, the quark and lepton transition
is possible in GUTs, triggering proton to decay. The lepto-quark
gauge bosons and colored scalars are responsible for proton decay.
The lepto-quark gauge boson mass is the unification scale $M_U$.
But the Higgs multiplet containing the SM Higgs doublet must
contain a light spectrum for the doublet. The colored partner of
the doublet must be superheavy, or proton lifetime is absurdly
short, $< 10^{-9}$ seconds, for O(1) couplings. 
For example, the ${\bf\bar 5}_H$ of
$SU(5)$ contains a Higgs doublet field $H_1(Y=-1/2)$ and color
triplet field $H_T(Y=1/3)$, where it is assumed that $H_1$ is
light and $H_T$ is superheavy. There exists the difficulty in
splitting the doublet-triplet masses, which is the split multiplet
problem.

Recently, it was pointed out that the split multiplet problem can
be understood in 5D theories with the $S_1/Z_2\times Z_2'$
orbifold compactification\cite{kawamura}. It is because of the
geometric twist of the gauge group such that some fields are
projected out from the massless spectrum. Indeed, the orbifold
compactification in string models\cite{dixon} has shown already
some models without colored scalars, realizing the split
multiplet.\footnote{The first model without colored scalars is
Model 3 of Ref.\cite{iknq}.} Thus, orbifold compactification in
higher dimensional theories may be the underlying reason for the
split multiplet\cite{kim,orbfermion,kimkim}. In the context of
this orbifold breaking of the GUT groups, some issues can be
reconsidered as for the gauge coupling
unification\cite{hn,hebecker,hns,thresh,running}, the larger GUT
groups\cite{so10}, and the flavor
unification\cite{bulksupp,running} In this paper, we try to
understand {\it geometrically} the top-bottom mass hierarchy and
the $s-\mu$ puzzle\cite{gj}.

The fifth dimensional coordinate $y$ is compactified to a torus
$2\pi R\equiv 0$. Furthermore, the point $y=-a$ is identified to
$y=a$ ($Z_2$ symmetry) and the point $y=(\pi R/2)+a$ is identified
to $y=(\pi R/2)-a$ ($Z_2'$ symmetry). This modding introduces a
fundamental region $y=[0, \pi R/2]$ and there arise two fixed
points, $y=0$ and $y=\pi R/2$. This geometry is used to twist the
GUT multiplet. In particular the GUT multiplet ${\bf\bar 5}_H$
living in the bulk is twisted, the twisting being represented by
$P= diag.(1,1,1,1,1)$ and $P'=diag.(-1,-1,-1,1,1)$. Obviously, the
twisting breaks $SU(5)$. But in the bulk the $SU(5)$ symmetry is
manifest above the unification scale and the gauge coupling
unification is assumed above $M_U$. The bulk fields are split into
four different Kaluza-Klein(KK) categories $\phi_{i,j}$ with the
$Z_2\times Z_2'$ quantum numbers $(i,j)$,
\begin{equation}
\phi_{++}=\sum^\infty_{n=0}a_{2n}\phi
^{(2n)}_{++}(x^\mu)\cos\frac{2ny}{R}
\end{equation}
\begin{equation}
\phi_{+-}=\sum^\infty_{n=0}a_{2n+1}\phi
^{(2n+1)}_{+-}(x^\mu)\cos\frac{(2n+1)y}{R}
\end{equation}
\begin{equation}
 \phi_{-+}=\sum^\infty_{n=0}a_{2n+1}\phi
^{(2n+1)}_{-+}(x^\mu)\sin\frac{(2n+1)y}{R}
\end{equation}
\begin{equation}
\phi_{--}=\sum^\infty_{n=0}a_{2n+2}\phi^{(2n+2)}_{--}(x^\mu)\sin\frac{(2n+2)y}{R}
\end{equation}
where $x^\mu$ is the 4D spacetime coordinate, $a_0=\sqrt{2/\pi R}$
and $a_n=\sqrt{4/\pi R}$ for $n\ne 0$. The massless field is
$\phi_{++}^{(2n)}$ for $n=0$. In this way, the massless 4D Higgs
doublet is obtainable from 5D while color triplets are all heavy.

Now let us extend the study to include $N=1$ supersymmetry. In 5D,
there exists an $N=2$ supersymmetry. One $Z_2$ breaks down the
$N=2$ down to $N=1$ and the other $Z_2$ breaks $G$ down to the SM.
Two 4D spinors(e.g. two Weyl spinors) make up one 5D spinor. Thus,
a 5D field thery is not anomalous. We can introduce only one
$SU(5)$ fermion multiplet in the bulk without worrying about the
anomaly, say a hypermultiplet ${\bf \bar 5}_H$. Upon
compactification, the $N=1$ supermultiplets are
\begin{eqnarray}
&H&_1^{(2n)}[(++);(1,2,-\frac{1}{2})],\ {\rm mass}=2n/R\\
&H&_T^{(2n+1)}[(+-);(3,1,\frac{1}{3})],\ {\rm mass}=(2n+1)/R\\
&\hat H&_T^{(2n+1)}[(-+);(\bar 3,1,-\frac{1}{3})],\ {\rm mass}=(2n+1)/R\\
&\hat H&_D^{(2n+2)}[(--);(1,2,\frac{1}{2})],\ {\rm mass}=(2n+2)/R
\end{eqnarray}
where the brackets [ ] contain the quantum numbers of $Z_2\times
Z_2'\times SU(3)\times SU(2)\times U(1)$. The original 5D $SU(5)$
theory with one anti-quintet is anomaly free. But the orbifolding
introduces one massless fermion doublet only, $H_1(n=0)$. The
other massive fields in the bulk pair up to form massive KK towers
of mass $m=n'/R$ where $n'=1,2,\cdots,\infty$. Since the low
energy theory should be anomaly free, we are dictated to introduce
brane fermions. So at one fixed point we introduce a 4D $N=1$ SM
Higgs supermultiplet $H_2$ with the quantum number
$(1,2,\frac{1}{2})$ under $SU(3)\times SU(2)\times U(1)$. At this
field theory level, the introduction of anomaly cancelling
fermions at the fixed points is arbitrary. It is needed from the
renormalizability of the low energy effective theory. However, the
orbifold compactification in string models introduces fixed point
fermions definitely once the bulk fermions carry
anomaly\cite{iknq,orbfermion}.

Under the framework of the preceding paragraph, we will consider
two models in Sec. II

\indent Model (I) One ${\bf\bar 5}_H$ in the bulk

\indent Model (II) One ${\bf\bar 5}_H$ plus $({\bf \bar
5}_{f,1}+{\bf \bar 5}_{f,2})$ in the bulk

\noindent where $H$ denotes a Higgs field and $f$ denotes some
fermions of the SM.

In Sec. III, we try to understand the $s-\mu$ puzzle geometrically
along the line of the split multiplet in the bulk, and present
Model (III) for an explicit presentation.

Since there appears the KK tower of the split multiplet in the
bulk we expect a correction to $\alpha_s(M_Z)$ from the usual SUSY
GUT prediction,
\begin{equation}
\delta\alpha_s(M_Z)\equiv\alpha_s^{exp}(M_Z)-\alpha_s^{SGUT,0}
\end{equation}
which is $\delta\alpha_s(M_Z)=-0.013\pm 0.0045$\cite{running}. The superscript 0
denotes no threshold correction. We will show that in Models (I),
(II) and (III) the Kaluza-Klein mode corrections are in the
favorable direction toward the experimental data.

\section{Splitting $H_1$ and $H_2$ in the bulk and at a brane}

At the minimal supersymmetric standard model(MSSM) level,
$H_1$(coupling to $b$ quark) and $H_2$(coupling to $t$ quark) are
not distinguished except for their gauge quantum numbers. Thus,
the apparent disparity of the top bottom masses is not understood.
It is fixed either by a large top Yukawa coupling and a small
bottom coupling with $\tan\beta\sim 1$ or by comparable Yukawa
couplings and a large $\tan\beta$. In this section, we explore a
possibility that the couplings and vacuum expectation values are
comparable, but the mass hierarchy is understood from a geometric
origin\cite{ah}. Namely, the origin of $H_1$ and $H_2$ are
different in a higher dimensional theory.\footnote{Without 
grand unification, separating $H_1$ and $H_2$ in the bulk
and brane was considered before\cite{tait}. However, in our
GUT theory, assigning $H_2$ at a brane is needed to explain the 
difference of $b-t$ mass scales in the low energy effective theory.}

To concentrate on the $b-t$ disparity, we restrict our discussion
to the third family only.

\subsection{Model (I)}

As the simplest model of the field theoretic orbifold
compactification, let us introduce a ${\bf\bar 5}_H$ in the 5D
bulk. The compactification is $S_1/Z_2\times Z_2'$ as shown in
Introduction. Because of the unification in the 5D bulk the gauge
coupling is unified above the GUT scale $M_U$ which can be a
string scale in a theory from string compactification. The
$S_1/Z_2\times Z_2'$ compactification produces one massless
supermultiplet $H_1$ containing one Higgs doublet. The
compactification is schematically drawn in Fig. 1 where two fixed
points(3-branes) $O$ and $A$ are shown and the thick line is the
fundamental region(=the bulk) in 5D. In the bulk $SU(5)$ gauge
fields and ${\bf\bar 5}_H$ live. At the 3-brane $A$ we locate the
missing Higgs doublet $H_2$ and the SM fields(including three
copies of supermultiplets of 15 chiral fields). The 5D Lagrangian
contains
\begin{eqnarray}
&S\supset \int d^4x\int_0^{\pi R/2}dy\left[
\partial^MH_1^\dagger (x,y)\partial_M H_1(x,y)+\delta(y-\frac{\pi R}{2})
(\lambda_bH_1QD^c+f_tH_2QU^c) \right]\\
&=\int d^4x\left[
\partial^\mu H_1^{(0)\dagger}(x)\partial_\mu H_1^{(0)}(x)
+y_bH_1^{(0)}Q_3D_3^c +y_tH_2Q_3U_3^c\right]
\end{eqnarray}
where $y_t=f_t,y_b=f_b\sqrt{2/\pi M_U R},\lambda_b=f_b/M_U^{1/2},$
and $H_1^{(0)}(x,y)=\sqrt{2/\pi R}H_1^{(0)}(x)$. Thus, we obtain
hierarchic masses
\begin{equation}
\frac{m_b}{m_t}=\frac{1}{\tan\beta\sqrt{M_UR\pi/2}}\sim
\frac{1}{60}.
\end{equation}

Note that the geometric suppression is the square root of $R$,
which may not be large enough. Therefore, to enhance the
suppression we consider the following model.

\subsection{Model (II)}

In Model (I) we inserted only ${\bf \bar 5}_H$ in the bulk. Here,
we introduce $b^c$ in the bulk also. The bulk field must be an
$SU(5)$ multiplet. For no mass hierarchy between $b$ and $\tau$ masses, 
we need a
complete multiplet. But an $SU(5)$ multiplet field in the bulk
allows only a split massless field. For a complete multiplet ${\bf
\bar 5}$ to be massless, we have to introduce two ${\bf \bar 5}$'s
so that an anti-quark singlet from one ${\bf \bar 5}$ and a lepton
doublet from the other ${\bf \bar 5}$ survives as a massless field
by appropriately twisting the bulk fields. In Fig. 2, we assign
the fields in the bulk and at the 3-brane $A$. Except the quintets
containing $b^c,\tau_L$, all the SM fermions are located at $A$.
Of course, we locate $H_2$ at $A$ to cancel the gauge anomaly. The
relevant 5D Lagrangian is
\begin{eqnarray}
&S\supset \int d^4x\int_0^{\pi R/2}dy\left[
\partial^MH_1^\dagger (x,y)\partial_M
H_1(x,y)+\bar D_3^c(x,y)i\partial_M\gamma^MD_3^c(x,y) +\bar
L_3(x,y)\right.\nonumber\\
&\left.\cdot i
\partial_M\gamma^ML_3(x,y)
+\delta(y-\frac{\pi R}{2}) (\lambda_bH_1Q_3D_3^c+f_tH_2Q_3U_3^c
+\lambda_\tau H_1L_3'E_3^c) \right]\\
&=\int d^4x\left[\cdots+y_bH_1^{(0)}Q_3D_3^{c(0)}
+y_tH_2Q_3U_3^c+y_\tau
H_1^{(0)}L_3^{\prime(0)}E_3^c\right]\nonumber
\end{eqnarray}
from which we obtain a linear relation in $R$, $y_b\sim
y_\tau\sim(\pi M_U R/2)^{-1}y_t$. Note that $\lambda_b\sim
f_b/M_U,\lambda_\tau\sim f_\tau/M_U, H_1^{(0)}(x,y)=\sqrt{2/\pi
R}H_1^{(0)}(x), D_3^{c(0)}(x,y)=\sqrt{2/\pi R}D^{c(0)}(x)$, and
$L_3^{\prime(0)}(x,y)=\sqrt{2/\pi R}L_3^{\prime(0)}(x)$.

\subsection{Running of gauge coupling constants}

The mass scales of interest in our scenario are the electroweak
scale, the unification scale(or the string scale) $M_U$, and the
inverse compactification length $M_c=1/R$. We assume that the
compactification mass is smaller than the unification mass so that
the running of the Kaluza-Klein(KK) towers between $M_U$ and $M_c$
helps toward the unification condition. Let us define the ratio of
these two scales as $2N$
\begin{equation}
N=\frac{M_U}{2M_c}.
\end{equation}
The masses of the KK modes are
\begin{eqnarray}
&(+,+):&\ \ 2nM_c\ \ \ \ \ \ \ \ \ \ b_i\ \ (b_i^0\ {\rm for\ }n=0)\nonumber\\
&(+,-):&\ \ (2n+1)M_c \ \ \ c_i\nonumber\\
&(-,+):&\ \  (2n+1)M_c\ \ \ \bar c_i\\
&(-,-):&\ \  (2n+2)M_c \ \ \ \bar b_i\nonumber
\end{eqnarray}
where the columns show $(P,P')$ quantum numbers, KK masses, and
the $\beta$ function coefficients. The tower of KK excitations up
to $M_U$ contributes to the running of gauge couplings at their
thresholds\cite{ddg}.

At the scale $\mu$ below the compactification scale, the gauge
coupling constant is
\begin{equation}
\frac{8\pi^2}{g_i^2(\mu)}=\frac{8\pi^2}{g^2_U} +b_i'
\ln\frac{M_c}{\mu}+b_i'\ln (2N)+(b_i+\bar b_i) \sum_{n=1}^{N}\ln
\frac{2N}{2n}+(c_i+\bar c_i)\sum_{n=1}^{N}\ln\frac{2N}{2n-1}
\end{equation}
up to a threshold correction $\Delta_i$. $g_U$ is the unification
coupling. Stirling's formula gives $\sum_{n=1}^{N}\ln(2N/2n)\simeq
N-\frac{1}{2}\ln(2\pi N)$ and
$\sum_{n=1}^{N}\ln\frac{2N}{2n-1}\simeq N-\frac{1}{2}\ln 2 \simeq
N.$ Thus, the low energy MSSM couplings become
\begin{equation}
\frac{8\pi^2}{g_i^2(\mu)}=\frac{8\pi^2}{g_U^2}+b_i'\ln\frac{M_c'}{\mu}
+\tilde
b'\ln\frac{M_U}{M_c'}+\frac{b}{2}\left[\frac{M_U}{M_c}-1\right]+\tilde
b_i\ln\frac{M_U}{M_c'}\label{master}
\end{equation}
where $M_c'=M_c/\pi$, and
\begin{eqnarray}
b&\equiv& b_i+c_i+\bar b_i+\bar c_i\ \ {\rm for\ all\ }i
 \nonumber\\
\tilde b_i&\equiv& b_i^0-\frac{1}{2}(b_i+\bar b_i)\nonumber\\
b_i'&=& (33/5,1,-3)\ \ {\rm in\  MSSM}\\
\tilde b'&=&b_i'-b_i^0\ {\rm for\ all\ }i.\nonumber
\end{eqnarray}
\vskip 0.5cm

\noindent {\bf Model (I):} In Model (I), from the fields ${\bf
\bar 5}_H$ in the bulk we obtain
\begin{equation}
b_i^H=2b_i^{H_0},\ \ \tilde b_i=b_i^{H_0}-\frac{1}{2}b_i^H=0,
\end{equation}
and from the field $H_2$ in the brane
\begin{equation}
\tilde b_i: (\tilde b_3, \tilde b_2, \tilde
b_1)^{H_2}=(0,\frac{1}{2},\frac{3}{10}).
\end{equation} From
the vector multiplet in the bulk, $ b^A_i=(2/3) b_i^{A_0}$,
\begin{equation}
\tilde b_i= b_i^{A_0}-\frac{1}{2} b_i^A: (\tilde b_3,\tilde
b_2,\tilde b_1)^V=(-6,-4,0).
\end{equation}
The sum of the brane Higgs and the bulk vector contributions
define the total value, $\tilde b_i=\tilde b_i^{H_2}+\tilde
b_i^V$. Therefore, from Eq. (\ref{master}), we obtain a relation
between couplings at the electroweak scale
\begin{equation}
\frac{1}{g_3^2}=\frac{12}{7}\frac{1}{g_2^2}-\frac{5}{7}\frac{1}{g_1^2}
+\frac{\tilde b}{8\pi^2}\ln\frac{M_U}{M_c'} \label{unify}
\end{equation}
where $\tilde b=\tilde b_3-(12/7)\tilde b_2+(5/7)\tilde b_1=3/14$.

Strong coupling unification considered in Ref.\cite{running} is
due to the duplication of matter fields appearing in the extension
of chiral multiplets to hypermultiplets which make the gauge
coupling be strong at high energy scales. But, in all models
considered here, most matter fields are living at the brane and
the 5-D gauge theory becomes asymptotically free: $b = -9,-7,-6$
in Eq. (\ref{master}) for Models I, II, and III, respectively.

In the unification models, such as in SUSY $SU(5)$, one can
determine the unification mass and gauge coupling constant
$\alpha_U$ at the unification scale. Namely, if $M_U$ and
$\alpha_U$ are given, one can predict $\alpha_i$ at the
electroweak scale. These coupling constants satisfies the relation
given in Eq.~(\ref{unify}), and the experimental values at $M_Z$
may not satisfy Eq.~(\ref{unify}). In our models, the onset of the
KK modes introduces another parameter $M_c$. Thus, we can satisfy
the condition (\ref{unify}) by appropriately choosing $M_c$.
However, in our split multiplet models the unified gauge coupling
constant is extremely small due to the asymptotic freedom. Namely,
$M_c$ turns out to be far below the string scale, $M_c'=4.5\times
10^9$~GeV with $M_U=2.8\times 10^{19}$~GeV.

However, it may be a better treatment of the problem if we satisfy
one condition. We choose the condition as the ratio $M_U/M_c'$ in
view of the experimental errors including the error in $\alpha_s$
and other effects (running effects due to Yukawa couplings and two
loop runnings). Thus, the unification condition (\ref{unify}) is
satisfied approximately but not exactly due to the error bars
allowed and hence we will study just the modification of
$\alpha_s$. Then, $\delta\alpha_s=\alpha_s^{exp}(M_Z)-
\alpha_s^{SGUT}$ where
$\alpha_s^{SGUT}=\alpha_s^{SGUT,0}+\Delta_{KK}\alpha_s$ is the KK
mode corrected value in \cite{running} with reasonable choices of
the ratio $M_U/M_c'$. For $M_U/M_c' = 10^2, 10^3, 10^4$, we obtain
$\Delta_{KK} \alpha_s \simeq -0.003, -0.004, -0.005$,
respectively. Then, $\delta\alpha_s=(-0.010, -0.009, -0.008)\pm
0.0045$ which correspond to $2.3\sigma, 2.0\sigma, 1.8 \sigma$
away from the experimental data. In all cases considered above the
additional logarithmic running reduces the discrepancy between
experimental value and the prediction of SUSY GUT even with no
threshold correction. The coupling at the unification scale
becomes $\alpha_{U}=2\times 10^{-2}, 4\times 10^{-3},$ and
$4\times 10^{-4}$, respectively. We used the conventional value $\alpha_s\simeq
1/24$ at $\mu=M_c$. At $M_U$ we also cutoff the power running and there
appears an O(1) uncertainty in $N$, which however 
does not affect the unification condition.

\vskip 0.5cm

\noindent {\bf Model (II):} We can repeat the same calculation for
Model (II). But note that the additional fields ${\bf\bar
5}_{f,1}$ and ${\bf\bar 5}_{f,2}$ have the following KK modes
\begin{eqnarray}
&{\bf\bar 5}_{f,1}=((++),(+-),(-+),(--))=(D_3^c,L_3,\hat L_3,\hat
D^c_3)\nonumber\\
&{\bf \bar 5}_{f,2}=((+-),(++),(--),(-+))= (D_3^{\prime
c},L_3',\hat L_3',\hat D_3^{\prime c})
\end{eqnarray}
so that the zero modes are $(D_3^c,L_3')_{n=0}$ which mimicks a
GUT multiplet. Another massive GUT-like multiplets are the even KK
modes $(++)=(D_3^c,L_3')_{n\ne 0}={\bf \bar 5}$ and $(--)=(\hat
D_3^c,\hat L_3')={\bf 5}$ which contribute to the log running. The
odd KK modes contributes only to the power running. Thus, the
difference of gauge couplings and the ratio $N$ are not changed,
viz. Eq.({\ref{unify}}).  For $\delta\alpha_s=0$ the unification
coupling is changed to $\alpha_{U} \simeq 1 \times 10^{-9}$ for
the exact unification. A similar analysis as in the study of Model
(I) for $M_U/M_c' = 10^2, 10^3,$ and $10^4$, the coupling constant at the
unification scale becomes $\alpha_{U} \simeq 2 \times 10^{-2},
5 \times 10^{-3},$ and $ 5 \times 10^{-4}$,
respectively. The KK mode correction to $\delta\alpha_s$ is the same
as in Model (I).

\section{Splitting the second family fermions in the bulk and at a brane}

As discussed in the preceding section, there are a lot of
possibilities for obtaining hierarchies of couplings by locating
some fields in the bulk and some fields at a brane. In this
section, we explore one more possibility for geometrically
generating hierarchical coupling structure. One of the puzzles in
the $SU(5)$ GUT is that in the second family the quark Yukawa
coupling is too small (by a factor of 3) compared to the lepton
coupling, which is the $s-\mu$ puzzle. To obtain a desired
suppression for the $s$ quark coupling, Georgi and Jarlskog
introduced ${\bf 45}_H$ in addition to the usual ${\bf
5}_H$\cite{gj}. In our scenario of keeping a split part of an
$SU(5)$ multiplet as a massless spectrum in the bulk, there is a
possibility of geometrically understanding the $s-\mu$ puzzle.

For simplicity, we modify the simplest example, Model (I) of the
previous section, and comment on another possibility after 
the discussion on Model (III). The Higgs fields, the first and 
the third family members are the same as in Model (I). We only 
change the members of the second family.

\centerline{\bf Model (III)}

Among the second family members, some fields are put in the bulk.
It is a split multiplet from ${\bf 10}$. In the bulk, the members
of ${\bf 10} \equiv (Q_2, U^c_2, E^c_2,\hat U^c_2,\hat E^c_2,\hat
Q_2) =[(3,2),(\bar 3,1),(1,1),(3,1),(1,1),(\bar 3,2)]$ under
$SU(3)\times SU(2)$ are assigned the $Z_2\times Z_2'$ parity as
$(++),(+-),(+-),(-+),(-+),$ and $(--)$, respectively. Thus, only
the quark doublet $Q_2$ has a zero mode spectrum $Q_2^{(0)}$ which
we interpret as the second family quark doublet. At low energy,
the theory must be anomaly-free and hence we locate the rest
members of the second family, $s^c,c^c,\mu^c,L_2=(\nu_\mu,\mu)_L$,
at the brane, which is shown in Fig. 3. The 5D Lagrangian contains
\begin{eqnarray}
&S\supset \int d^4x\int^{\pi R/2}_0\left[\partial^M
H_1^\dagger(x,y)
\partial_MH_1(x,y)+\bar Q_2(x,y)i\partial_M\gamma^MQ_2(x,y)+\delta(y-
\frac{\pi R}{2})(\lambda_sH_1Q_2D_2^c \right.\nonumber\\
&+\left. \lambda_cH_2Q_2U_2^c+\lambda_\mu
H_1L_2\mu^c+\lambda_bH_1Q_3D_3^c+f_tH_2Q_3U_3^c+\cdots)\right]\\
&=\int d^4x\left[y_sH_1^{(0)}Q_2^{(0)}D_2^c+y_\mu
H_1^{(0)}L_2E_2^c+\cdots\right]\nonumber
\end{eqnarray}
Note that $\lambda_s\sim f_s/M_U, \lambda_\mu\sim
f_\mu/\sqrt{M_U}, y_s=(2/\pi M_UR)f_s, y_\mu=\sqrt{2/\pi M_U
R}f_\mu$, implying
\begin{equation}
\frac{y_s}{y_\mu}\sim\frac{1}{\sqrt{M_UR}}
\end{equation}
Thus, the strange quark Yukawa coupling is suppressed compared to
the muon Yukawa coupling.

\indent From the bulk zero mode $Q_2^{(0)}$, $\tilde
b_i=b_{Q_2}^{(0)}-(1/2)(b_{Q_2}+b_{\hat Q_2})=0$. From the brane
fields, $U_2^c$ and $E_2^c$, $\tilde b_i=(b_{U^c})_i+(b_{E^c})_i
=(1/2,0,4/5)+(0,0,3/5)=(1/2,0,7/5)$ for $i=3,2,1$. From the brane
Higgs $H_2$, $\tilde b_i=(0,1/2,3/10)$. From the bulk Higgs $\bar
5_{H_1} $, $\tilde b_i=0$. From the vector multiplet, $\tilde
b_i=(-6,-4,0)$. Thus, we obtain
\begin{equation}
\tilde b_i=(-\frac{11}{2},-\frac{7}{2},\frac{17}{10})\ \ {\rm for\
} i=3,2,1.
\end{equation}
Therefore,
\begin{equation}
\tilde b=\tilde b_3-\frac{12}{7}\tilde b_2+\frac{5}{7}\tilde
b_1=\frac{12}{7}.
\end{equation}

Model (III) is interesting since it turns out that $M_c'$ is very
large $\simeq 1.5 \times 10^{15} {\rm GeV}$ and $M_U$ is the usual
unification scale $2.5\times 10^{16}$~GeV in order to satisfy the
Eq. (\ref{unify}). The unification coupling constant is also close
to the SGUT value $\alpha_{U} \simeq 0.03 \simeq \frac{1}{30}$.
Due to the large $\tilde{b}$ compared to the other models, Model
(III) allows the perfect unification with the logarithmic running
between a small scale difference of $M_U/M_c'=10^{1.22}$.

If we consider Model (III) with the backbone of Model (II) instead
the backbone of Model (I), we obtain a better relation between the
top-bottom mass hierarchy since the suppression will be linear in
$1/R$. Assuming all the couplings to be order 1,
we obtain $m_\mu/m_s\sim \sqrt{m_t/m_b\cdot\tan\beta}=\sqrt{60/\tan\beta}
\sim 3$ at the unification scale, implying $\tan\beta\sim 7$. 
$\delta\alpha_s$ is the same as in Model (III).

\section{Conclusion}

In this paper, we studied a new possibility for a field theoretic
orbifold compactification possessing supersymmetry, which was
applied to the top-bottom mass hierarchy and the $s-\mu$ puzzle.
This possibility relies on the missing massless spectrum in the
bulk. The 5D bulk theory with any fermion representation is
anomaly free, but the orbifold compactification may project out
split multiplet in the bulk. This situation has been observed in
orbifold compactifications in string models\cite{iknq,orbfermion}.
Because of the split multiplet, the anomaly-free condition
dictates to put some massless fermions at the brane so that the
resulting 4D theory is anomaly free. In string examples, the field
content and assignment of the fields at the fixed points are
determined uniquely by the modular invariance requirement. But in
our field theoretic example, the field content and the location
are arbitrary. In this paper, we chose the simplest possibility.

In our examples, we put the ${\bf \bar 5}_H$(containing the $H_1$
Higgs doublet) in the bulk. Because of the twisting, only $H_1$
from the ${\bf\bar 5}_H$ remains massless in the bulk. Thus, the
needed $H_2$ is put at a brane where the SM fields are located.
Thus, the bottom and the top quarks have geometrically different
factors for the effective 4D Yukawa couplings, rendering a
top-bottom mass hierarchy. To enhance the hierarchical factor some
SM fermions are put in the bulk in our second example. Similarly,
the $s-\mu$ puzzle is understood geometrically by putting the
strange quark doublet in the bulk, thus reducing the strange quark
Yukawa coupling compared to the muon Yukawa coupling. There are
other applications along this line, e.g. reducing the up quark
mass compared to the down quark mass. In all these examples we
considered, the corrections to the strong coupling constant are in
the right direction, making the low energy effective MSSM
predictions closer to the experimental value.

One tempting question to ask in this scenario might be the $\mu$
problem\cite{mu}. Certainly, one cannot write $\mu H_1H_2$ in the
bulk. It can be written only at the brane A. But the need to
introduce $H_2$ at A is below the compactification scale
$M_c=1/R$. Therefore, writing the dimensional parameter such as
$\mu$ must have a suppression factor, certainly less than $M_c$.
But at this moment, we do not understand geometrically how large
the suppression factor is. We may need an additional discrete or
global symmetry to sufficiently suppress $\mu$. In any case, these
extra symmetries are needed for proton longevity.

The field theoretic orbifolding considered recently is very simple
compared to the string theory orbifolding. However, it seems to be
arbitrary in choosing and assigning the fields, and we expect that
some string compactification in the future may lead to the above
types of field theoretic orbifolding so that the assignment of the
$H_2$ and the SM fields at the brane is no longer arbitrary.
\vskip 0.5cm

{\bf Note added:} Recently, there appeared an 
argument\cite{zwirner} that it
would not be possible to have a consistent SUSY field theory on the
$S^1/(Z_2\times Z_2^\prime)$ orbifold with a single bulk Higgs
multiplet, since there are gauge anomalies localized at the orbifold 
fixed points\cite{anomaly}. In our case, however, the local gauge anomalies can
be cancelled by introducing a brane Higgs field and 5D Chern-Simons terms
in the bulk\cite{callan,anomaly}. The models considered in the
present paper are consistent up to introducing the Chern-Simons terms.

\acknowledgments  This work is supported in part by the BK21
program of Ministry of Education, the KOSEF Sundo Grant, and by
the Center for High Energy Physics(CHEP), Kyungpook National
University.

\vskip 0.3cm
\begin{figure}[b]
\centering \centerline{\epsfig{file=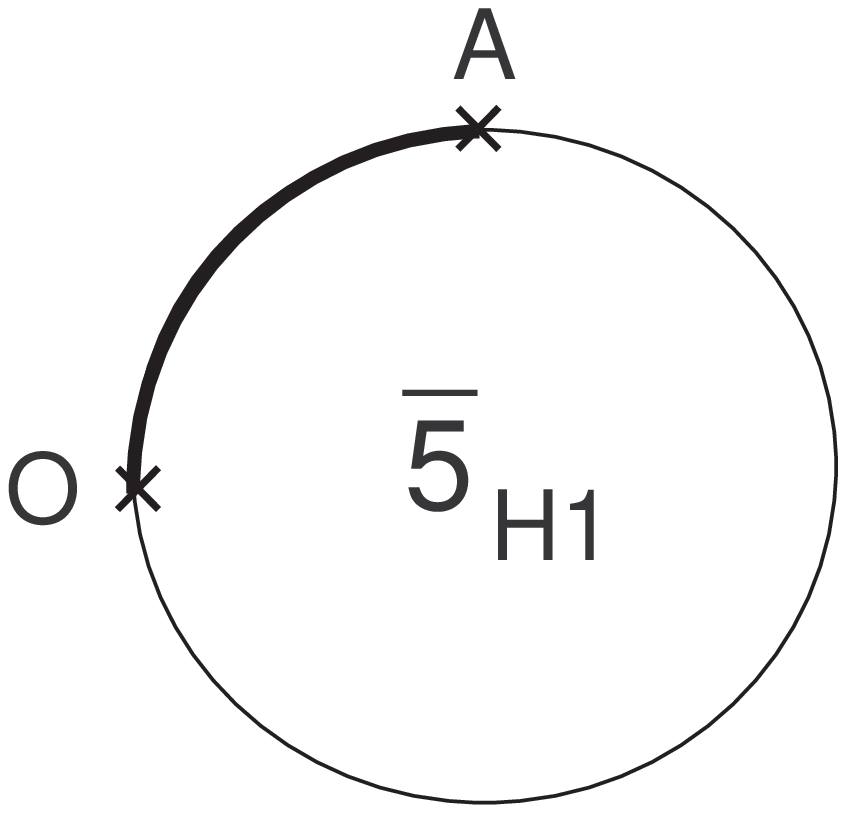,width=54mm}}
\end{figure}
\centerline{ Fig.~1.\ \it  ${\bf \bar 5}_H$ is put in the bulk,
and  the SM fields and $H_2$ are located at the fixed point A.}
\vskip 0.3cm

\vskip 0.3cm
\begin{figure}[b]
\centering \centerline{\epsfig{file=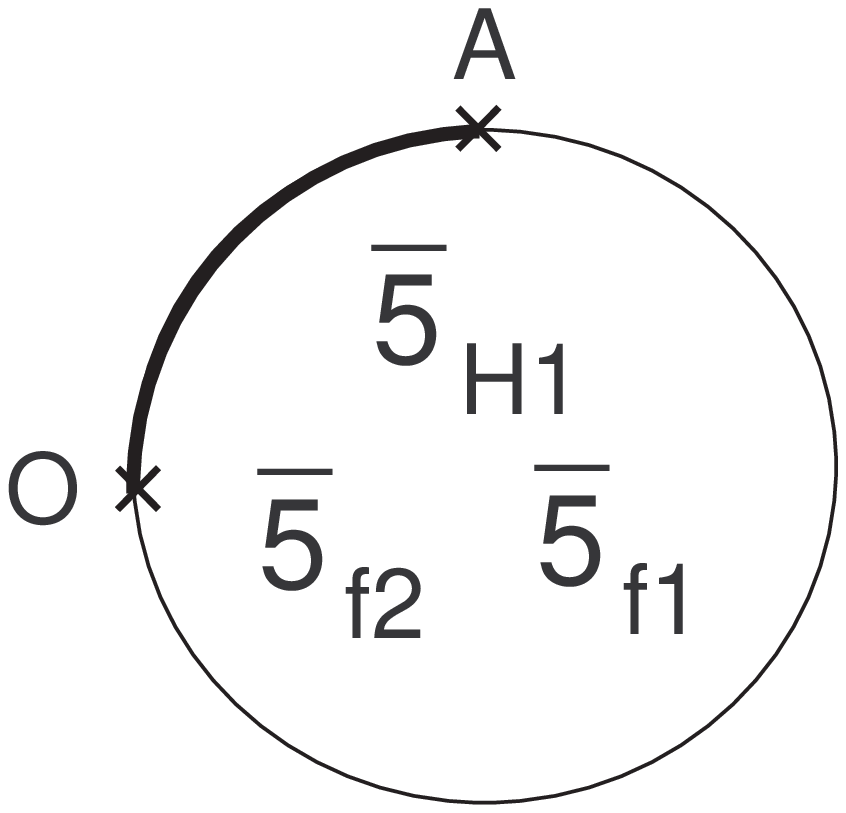,width=60mm}}
\end{figure}
\centerline{ Fig.~2.\ \it Same as Fig. 1 except that the
$(\nu_\tau,\tau)$ doublet and $b^c$ are put in the bulk.}

\vskip 0.3cm
\begin{figure}[b]
\centering \centerline{\epsfig{file=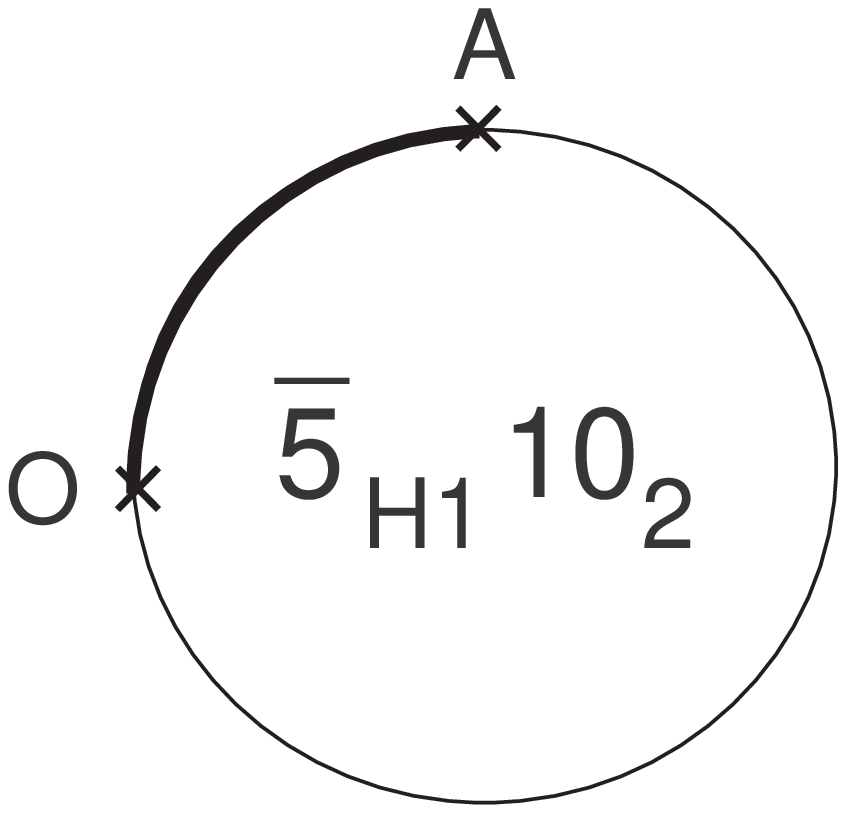,width=60mm}}
\end{figure}
\centerline{ Fig.~3.\ \it Same as Fig. 1 except that the $(c,s)$
doublet is put in the bulk.} \vskip 0.3cm

\end{document}